\documentstyle[epsfig]{mn2e}

\begin{document}
\def\ltsima{$\; \buildrel < \over \sim \;$}
\def\simlt{\lower.5ex\hbox{\ltsima}}
\def\gtsima{$\; \buildrel > \over \sim \;$}
\def\simgt{\lower.5ex\hbox{\gtsima}}
\def\approxgt{\mathrel{\hbox{\rlap{\lower.55ex \hbox {$\sim$}}
        \kern-.3em \raise.4ex \hbox{$>$}}}}
\def\approxlt{\mathrel{\hbox{\rlap{\lower.55ex \hbox {$\sim$}}
        \kern-.3em \raise.4ex \hbox{$<$}}}}

\title[On the origin of soft X-rays in obscured AGN]
{On the origin of soft X-rays in obscured AGN: answers
from high-resolution
spectroscopy with XMM-Newton}

\author[Matteo Guainazzi \& Stefano Bianchi]
{Matteo Guainazzi, Stefano Bianchi\\ ~ \\
European Space Astronomy Center of ESA, Apartado
50727, E-28080 Madrid, Spain\\
}

\maketitle
\begin{abstract}

We present results of a high-resolution soft X-ray (0.2--2~keV)
spectroscopic study of a sample of 69 nearby obscured
Active Galactic Nuclei (AGN)
observed with the Reflection Grating Spectrometer (RGS)
on board XMM-Newton. This is the largest sample ever studied
with this technique so far. The main conclusions of our study
can be summarized as follows: a) narrow
Radiative
Recombination Continua are detected
in about 36\% of the objects in
our sample (in 26\% their intrinsic width is
$\le 10$~eV); b) higher order transitions are
generally enhanced with respect to 
pure photoionization, indicating that resonant scattering
plays an important role in the ionization/excitation balance.
These results support the scenario,
whereby the active nucleus is responsible for the 
X-ray ``soft excess''
almost ubiquitously observed in nearby obscured AGN via
photoionization of circumnuclear gas.
They confirm on a statistical basis the conclusions
drawn from the detailed study of the brightest
spectra in the sample.
Furthermore, we propose a criterion to
statistically discriminate between
AGN-photoionized sources and starburst galaxies, based on
intensity of the forbidden component of the
O{\sc vii} He-$\alpha$ triplet (once normalized
to the O{\sc viii} Ly-$\alpha$)
coupled with the integrated luminosity in
He-like and H-like oxygen lines.

\end{abstract}


\section{Introduction}

Nearby X-ray obscured Active Galactic Nuclei (AGN) invariably exhibit 
excess emission above the extrapolation of the absorbed nuclear
emission \cite{turner97,guainazzi05}. The origin of this
component - which can represent a significant fraction of
the active nucleus bolometric energy budget 
\cite{levenson02} - is still largely unknown.
Gas in the nuclear environment could
be heated to million degrees
by shocks induced by AGN outflows
\cite{king05} or episodes of intense
star formation \cite{cidfernandes98,gonzalezdelgado01}.
Alternatively, the AGN primary emission
could photoionize and photoexcite circumnuclear gas.

The latter
scenario has recently received direct
observational support, thanks to high-resolution capabilities
in the spatial and frequency domains that large X-ray
observatories such as {\it Chandra} and XMM-Newton nowadays offer.
High-resolution spectra unveiled
signatures of photoionized plasma in a few bright objects:
NGC~1068 \cite{kinkhabwala02,young01}, the
Circinus Galaxy \cite{sambruna01}, Mkn~3 \cite{sako00,bianchi05,pounds05};
NGC~4151 \cite{schurch04}.
In NGC~1068 the contribution of collisionally ionized plasma
to the observed soft X-ray emission is constrained to be lower then
10\% \cite{brinkman02}.
These conclusions are based on three pieces of
experimental evidence:

\begin{enumerate}

\item the spectra are dominated by strong emission lines
of highly-ionized species from Carbon to Silicon,
as well as by L-shell transitions from Fe{\sc xvii} to
Fe{\sc xxi}

\item narrow Radiative Recombination Continua (RRC)
features from Carbon and Oxygen were detected.
The width of
these features indicates typical plasma temperatures of
the order of a few eV \cite{kinkhabwala02}.
These
features are unequivocal signatures of photoionized
spectra \cite{liedahl96}.

\item the intensity of higher-order series emission
lines,
once normalized to the K$_{\alpha}$, are larger than predicted
by pure photoionization, and are consistent with an important
contribution by
photoexcitation (resonant scattering) \cite{band90,matt94,krolik95}. This explains why
standard plasma diagnostics \cite{porquet00} fail
to properly interpret
the physical nature of the spectra.

\end{enumerate}

A solution in terms of AGN-photoionized
gas can also explain the coincidence in extension
and overall morphology between soft X-ray 
emission and the
Narrow Line Regions (NLRs), the latter traced by O[{\sc iii}] HST maps,
on scales as large as a few hundred parsecs
\cite{bianchi06}. Solutions satisfying the observed
X-ray to optical flux ratio require an approximately constant
ionization parameter ({\it i.e.}, a density scaling as the inverse
square of the distance from the ionizing source), similarly
to what is often found using photoionization models of the NLRs.

So far, high-resolution X-ray spectra have been published
only for a few X-ray bright obscured AGN.
However, diagnostically
important emission lines in these objects
exhibit very large
Equivalent Widths, $EW$s, as the
continuum is often totally suppressed in the
soft X-ray band. In
this {\it paper} we present the first systematic
high-resolution X-ray spectroscopic
study on a sizable sample of obscured AGN.
The Reflection
Grating Spectrometer (RGS) on board XMM-Newton
\cite{derherder01} is the most suitable instrument
currently flying for this purpose,
due to its unprecedented effective
area in the 0.2-2~keV band, as well as its good
absolute aspect solution accuracy ($\simeq$8~m\AA).

\section{{\it CIELO-AGN}}

\subsection{The sample}

Our sample comprises all the type $\ge$1.5 AGN
(according to the NED classification)
observed by XMM-Newton, and whose data were public
as of September 2006 (Tab.~\ref{tabidl6}
shows the list of sources in the
sample, together
with their redshift and the combined exposure time
for one RGS camera).
%----------------- Table IDL 6
\begin{table}
\caption{{\it CIELO-AGN} sources.}
\begin{tiny}
\begin{center}
\begin{tabular}{lcc} \hline \hline
Source & $z$   & Exposure time  \\
       &       & (ks)           \\
Circinus~Galaxy       & 0.001 & 102.2  \\
ESO509-G66           & 0.045 &  13.6  \\
IC2560               & 0.010 &  81.4  \\
IC4395               & 0.036 &  22.1  \\
IC4995               & 0.016 &  11.5 \\
IIIZW035             & 0.027 &  20.7 \\
IRAS01475-0740       & 0.018 &  11.5 \\
IRAS08572+3915       & 0.058 &  28.6 \\
IRAS09104+4109       & 0.442 &  13.5 \\
IRAS10214+4724       & 2.286 &  53.4 \\
IRAS13197-1627       & 0.017 &  44.6 \\
IRAS15480-0344       & 0.030 &  10.9 \\
MCG-5-23-16          & 0.009 &  61.6 \\
MRK3           & 0.014 & 180.8 \\
MRK6           & 0.019 &  96.4 \\
MRK331               & 0.019 &  17.9 \\
MRK348               & 0.015 &  46.3 \\
MRK612               & 0.020 &  11.8 \\
MRK744               & 0.009 &  22.7 \\
MRK993               & 0.015 &  22.6 \\
MRK1152              & 0.053 &  26.4 \\
NGC1068              & 0.004 &  86.8 \\
NGC1365              & 0.005 & 127.5 \\
NGC1386              & 0.003 &  16.9 \\
NGC1410              & 0.025 &  11.7 \\
NGC1614              & 0.016 &  23.5 \\
NGC2110              & 0.008 &  51.7 \\
NGC2273              & 0.006 &  12.5 \\
NGC2623              & 0.019 &  12.3 \\
NGC2992              & 0.008 &  28.5 \\
NGC34                & 0.020 &  21.8 \\
NGC3982              & 0.004 &  34.0 \\
NGC4138              & 0.003 &  14.3 \\
NGC4151              & 0.003 &  55.5 \\
NGC4168              & 0.007 &  23.0 \\
NGC424               & 0.012 &   7.9 \\
NGC4258              & 0.002 &  78.5 \\
NGC4303              & 0.005 &  42.8 \\
NGC4395              & 0.001 & 108.4 \\
NGC4472              & 0.003 &   1.5 \\
NGC4477              & 0.004 &  13.3 \\
NGC449               & 0.016 &  11.9 \\
NGC4507              & 0.012 &  44.5 \\
NGC4565              & 0.004 &  14.4 \\
NGC4639              & 0.003 &  14.6 \\
NGC4725              & 0.004 &  17.6 \\
NGC4945              & 0.002 &  63.8 \\
NGC4968              & 0.010 &  11.8 \\
NGC5033              & 0.003 &  19.7 \\
NGC5252              & 0.023 &  65.8 \\
NGC526A              & 0.019 &  46.9 \\
NGC5273              & 0.004 &  16.1 \\
NGC5506              & 0.006 &  32.9 \\
NGC5643              & 0.004 &   9.6 \\
NGC591               & 0.015 &  11.8 \\
NGC6552              & 0.026 &   6.8 \\
NGC7172              & 0.009 &  72.0 \\
NGC7212              & 0.027 &  13.7 \\
NGC7314              & 0.005 &  43.4 \\
NGC7479              & 0.008 &  12.6 \\
NGC7582              & 0.005 &  22.3 \\
NGC7674              & 0.029 &  10.3 \\
UGC1214              & 0.017 &  11.8 \\
UGC2456              & 0.012 &  17.1 \\
UGC2608              & 0.023 &   7.5 \\
UGC4203              & 0.014 &   7.8 \\
UGC6527              & 0.027 &  23.6 \\
UGC8621              & 0.020 &  11.8 \\
UM625                & 0.025 &  11.6 \\
\hline \hline
\end{tabular}
\end{center}
\end{tiny}
\label{tabidl6}
\end{table}
%----------------- Table IDL 6
After excluding
H1320+551 and H1419+480 due to uncertainties in
their classification, the sample
includes 69 sources.

\subsection{Data processing}

For each
observation, we have reprocessed RGS data starting
with the {\it Observation Data Files}, using
SASv6.5 \cite{gabriel03}, and the most advanced
calibration files available as of May 2006.
Background spectra were generated using blank field
event lists, accumulated from different positions on
the sky vault along the mission.
Spectra of the same source from different observations
were merged, together with their response
matrices, after checking that no significant
spectral variability occurred.

Each spectrum was
systematically searched for the presence of
emission lines. We simultaneously fit 
together the spectra of the two RGS cameras, using
{\sc Xspec} version 12.3.0.
``Local'' fits to the data were
performed on the unbinned spectra on $\simeq$100~channels wide
intervals, using
Gaussian profiles to account for
any line features. In only a few cases (see Tab.~\ref{tab2})
%------------------ Tab.2
\begin{table}
\caption{1-$\sigma$ intrinsic widths for
resolved Gaussian
profiles in {\it CIELO-AGN}.}
\begin{center}
\begin{tabular}{lc} \hline \hline
Transition & Width \\
 & (km~s$^{-1}$) \\ \hline
\multicolumn{2}{l}{NGC1068} \\
Fe{\sc xvii} 3d-2p ($^1P_1$) & $670 \pm^{110}_{10}$ \\
N{\sc vii} Ly-${\alpha}$ & $820 \pm^{80}_{40}$ \\
O{\sc vii} He-${\alpha}$ & $550 \pm^{50}_{20}$ \\
O{\sc vii} He-${\beta}$ & $3300 \pm^{200}_{600}$ \\
O{\sc vii} He-${\gamma}$ & $650 \pm^{150}_{50}$ \\
O{\sc viii} Ly-${\alpha}$ & $460 \pm^{60}_{40}$ \\
O{\sc viii} Ly-${\beta}$ & $1000 \pm^{200}_{20}$ \\ \hline
\multicolumn{2}{l}{NGC4151} \\
Fe{\sc xviii} 3s-2p & $1100 \pm^{900}_{600}$ \\
C{\sc vi} Ly-${\alpha}$ & $490 \pm^{150}_{80}$ \\
C{\sc vi} Ly-${\beta}$ & $900 \pm^{700}_{400}$ \\ \hline
\multicolumn{2}{l}{NGC4507} \\
O{\sc vii} He-${\alpha}$ & $520 \pm^{110}_{190}$ \\ \hline \hline
\end{tabular}
\end{center}
\label{tab2}
\end{table}
%------------------ Tab.2
the width of the profiles accounting for
bound-bound transitions is inconsistent
with zero; hence in these cases only
the intrinsic profile width has been left free
in the fit. On the other hand,
free-bound transitions were
modeled with Gaussian profiles, where the intrinsic
width was always considered an additional
free fit parameter. We have always assumed
the same intrinsic width for the components of a
multiplet. Local
continua were modeled with
$\Gamma = 1$ power-laws, leaving free in each
fit the continuum normalization ($\Gamma$ is
the power-law photon index). No assumption was
made {\it a priori} on the line centroid energies.\footnote{On the
other hand, upper limits on the line intensities
were calculated assuming the laboratory energies.}
However, fits of He-$\alpha$ triplets have been
performed keeping the relative distance between the
centroid energies of the components fixed to the value
dictated by atomic physics.
A line is considered to be detected when its flux
is inconsistent with 0 at the 1-$\sigma$ level.
In Appendix~A we list energies, fluxes and
intrinsic widths (for the RRC only) of the transitions
individually presented in this paper, and discuss the consistency
of our measurements with the expected laboratory energies
[laboratory energies are extracted
from the {\sc CHIANTI} database \cite{dere01}].
Line luminosities have been corrected for Galactic
photoelectric absorption using column densities after
Dickey \& Lockman (1990).

We have as well checked whether
the detected line energy centroids in each source
were systematically shifted with respect to
the laboratory energies. We define ``systematic''
energy shifts
different from zero (in the
same direction) at the 1$\sigma$ level
in at least
two of the following
transitions: C{\sc vi} Ly-${\alpha}$,
O{\sc vii} He-${\alpha}$,
O{\sc vii} He-${\beta}$, and O{\sc viii} Ly-${\alpha}$.
Our analysis found none of these systematic shifts.
Nonetheless, we found some
occurrences of discrepancies between the
best-fit centroid and the laboratory energies in
individual lines. We refrain from attributing any
astrophysical meaning to these discrepancies, which
may be due to contamination by nearby
transitions, residual errors in the aspect solution
of the RGS cameras, or spurious detections. The reader
may find a more detailed discussion of this point in
the Appendix.

In Fig.~\ref{fig1} we show the RGS spectra of the three
%---------------------------------------
\begin{figure*}
\hbox{
\epsfig{file=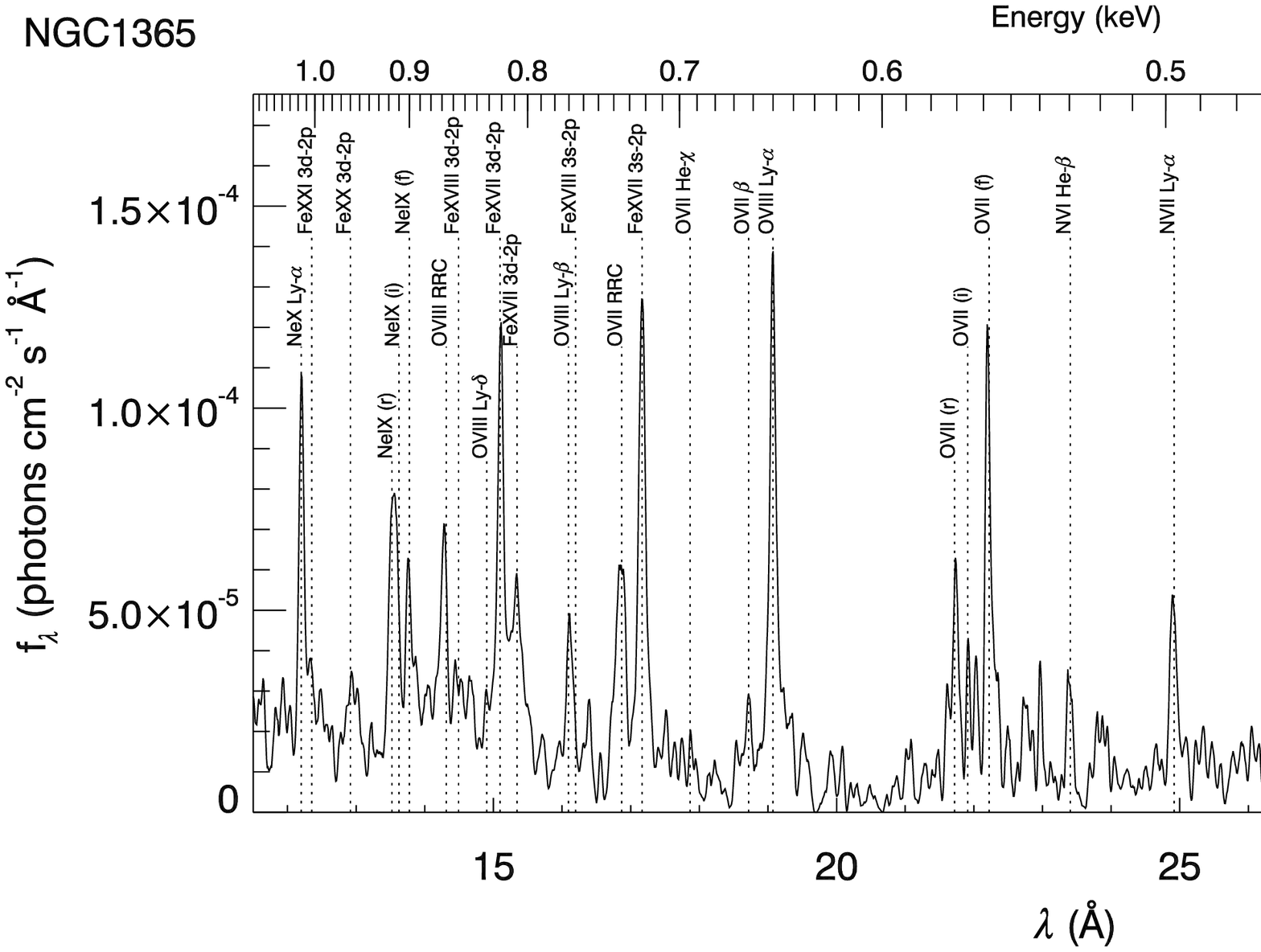,height=72mm,width=160mm}
}
\hbox{
\epsfig{file=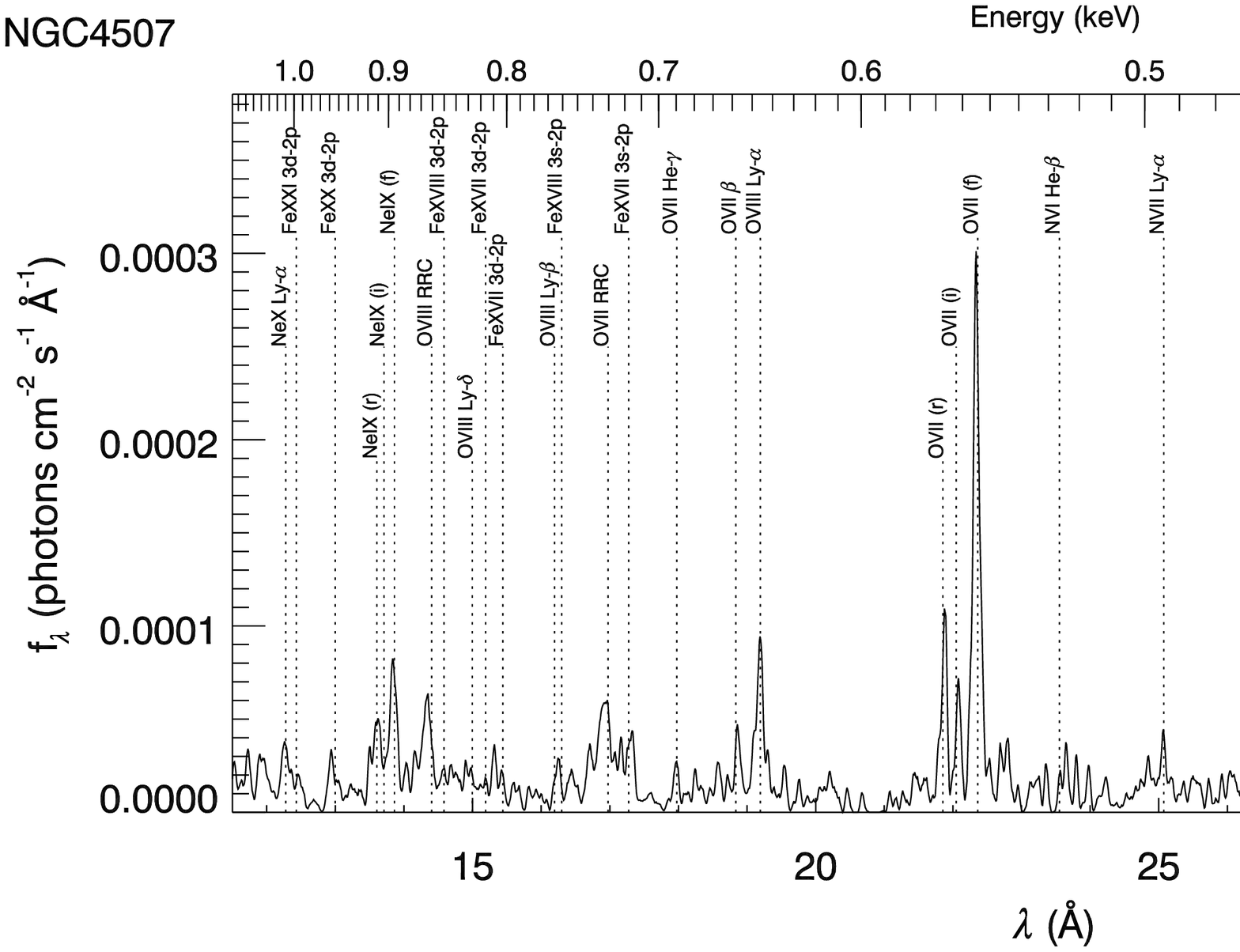,height=72mm,width=160mm}
}
\hbox{
\epsfig{file=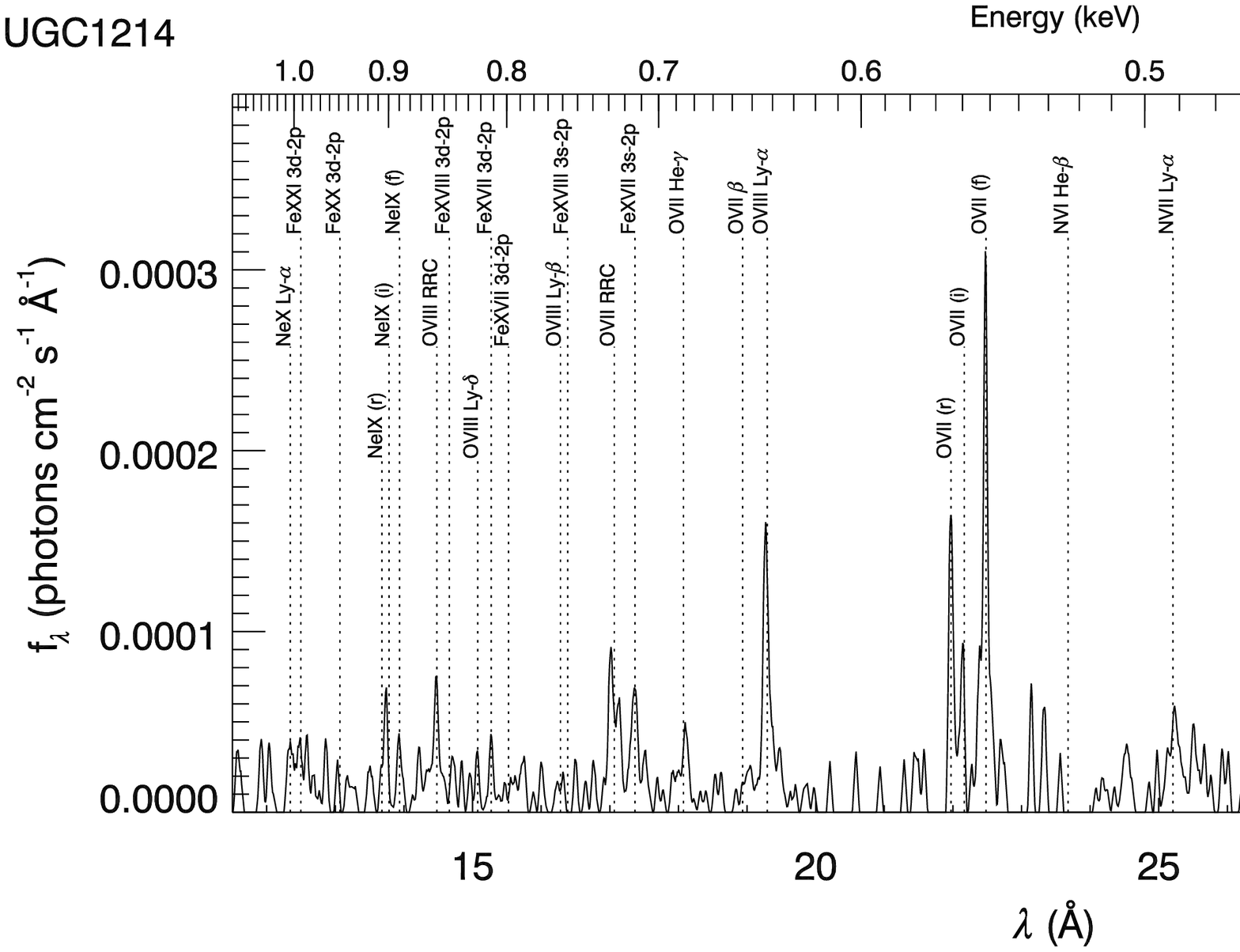,height=72mm,width=160mm}
}
\caption{RGS spectra for the three galaxies of our sample,
which exhibit the largest number of detected emission lines
(in brackets below) and whose RGS spectra have not
been published yet: NGC~1365 (19),
NGC~4507 (17), UGC~1214 (14)
[for comparison
the number of detected lines in the brightest
Seyfert~2 of our samples are:
25 (NGC~1068), 19 (the Circinus Galaxy),
18 (NGC~4151), 17 (Mrk~3)]. Spectra
of the two RGS cameras have been merged and
smoothed with a 5-channels wide triangular kernel
for illustration purposes only. The positions
of the line transitions
measured in {\it CIELO-AGN} are labeled.}
\label{fig1}
\end{figure*}
%---------------------------------------
still unpublished Seyfert~2 galaxies in our sample,
which exhibit the
largest number of line detections.

Hereafter,
uncertainties on the fitting parameters
are quoted at the 1$\sigma$ level; likewise the upper limits
represent a 1$\sigma$ confidence level.
Errors on the line centroid energies
include a 8~m\AA systematic error.
Throughout this paper,
energies/wavelengths of astrophysical lines
are quoted in the source rest frame, unless
otherwise specified.

\subsection{The catalog}

The results of the procedure outlined above have been
compiled into {\bf CIELO-AGN} ({\it Catalog
of Ionized Emission Lines in Obscured AGN}).
Our study exhibit a
high-detection efficiency, at least
with respect to the most common lines
observed in obscured AGN
spectra, despite the overall
low soft X-ray flux (sample median
6$\times$10$^{-13}$~erg~cm$^{-2}$~s$^{-1}$; see Tab.~\ref{tabidl1}
for a complete list of the lines detected in {\it CIELO-AGN}).
%--------------- Table IDL 1
\begin{table} 
\caption{Number of {\it CIELO-AGN} sources, $N_{det}$, for which                            
a given {\it Transition} (laboratory wavelength                                 
$\lambda_{lab}$) has been detected, if $N_{det}>0$}       
\begin{center}                                                            
\begin{tabular}{lcc} \hline \hline                                              
Transition & $\lambda_{lab}$ & $N_{det}$ \\                                     
           & ($\AA$)         &           \\ \hline                              
Si{\sc xiii} He-$\alpha$       &  6.740 &  5 \\
Mg{\sc xi} He-$\alpha$ ($r$)   &  9.228 & 12 \\
Mg{\sc xi} He-$\alpha$ ($f$)   &  9.314 & 13 \\
Ne{\sc x} Ly-${\alpha}$        & 12.134$^a$ & 20 \\
Fe{\sc xxi} 3d-2p              & 12.282 & 21 \\
Fe{\sc xx} 3d-2p               & 12.845 & 19 \\
Ne{\sc ix} He-$\alpha$ ($r$)   & 13.447 & 13 \\
Ne{\sc ix} He-$\alpha$ ($i$)   & 13.553 &  6 \\
Ne{\sc ix} He-$\alpha$ ($f$)   & 13.699 & 15 \\
O{\sc viii} RRC                & 14.228 & 13 \\
Fe{\sc xviii} 3d-2p            & 14.413$^b$ & 24 \\
O{\sc viii} Ly-$\delta$        & 14.821$^c$ & 20 \\
Fe{\sc xvii} 3d-2p ($^1P_1$)   & 15.015 & 25 \\
Fe{\sc xvii} 3d-2p ($^3D_1$)   & 15.262 & 24 \\
O{\sc viii} Ly-$\beta$         & 16.006 & 22 \\
Fe{\sc xviii} 3s-2p            & 16.091 & 23 \\
O{\sc vii} RRC                 & 16.771 & 17 \\
Fe{\sc xvii} 3s-2p (3G/M2)     & 17.076$^d$ & 27 \\
O{\sc vii} He-$\gamma$         & 17.768$^e$ & 16 \\
O{\sc vii} He-$\beta$          & 18.627 & 20 \\
O{\sc viii} Ly-$\alpha$        & 18.967 & 30 \\
O{\sc vii} He-$\alpha$ ($r$)   & 21.602 & 22 \\
O{\sc vii} He-$\alpha$ ($i$)   & 21.803 & 13 \\
O{\sc vii} He-$\alpha$ ($f$)   & 22.101 & 23 \\
Ne{\sc vii} Ly-$\alpha$        & 24.782$^f$ & 24 \\
C{\sc vi} Ly-${\beta}$         & 28.459$^g$ & 14 \\
N{\sc vi} He-$\alpha$          & 28.787 & 11 \\
C{\sc v} RRC                   & 31.622 &  6 \\
C{\sc vi} Ly-$\alpha$          & 33.737$^h$ & 23 \\
\hline \hline                                                                   
\end{tabular}  
\end{center}
                                                                 
\noindent                                                                       
$^a$doublet: $\lambda_1=12.1321$~\AA, $\lambda_2=12.1375$~\AA    
               
\noindent                                                                       
$^b$triplet: $\lambda_1=14.3760$~\AA, $\lambda_2=14.4187$~\AA,                  
$\lambda_3=14.4210$~\AA                  
                                       
\noindent                                                                       
$^c$possible contamination by Fe{\sc xx} 3p$^2$-2p$^2$ at $\lambda=14.8501$~\AA 
 
\noindent                                                                       
$^d$doublet: $\lambda_1=17.0500$~\AA, $\lambda_2=17.0970$~\AA  
                 
\noindent                                                                       
$^e$possible contamination by Fe{\sc xviii} 3p$^2-$2p$^2$ at                     
$\lambda=17.8472$~\AA                         
                                  
\noindent                                                                       
$^f$doublet: $\lambda_1=24.7790$~\AA, $\lambda_2=24.7840$~\AA    
               
\noindent                                                                       
$^g$doublet: $\lambda_1=28.4612$~\AA, $\lambda_2=28.4662$~\AA 
                  
\noindent                                                                       
$^h$doublet: $\lambda_1=33.7342$~\AA, $\lambda_2=33.7396$~\AA 
                                                        
\label{tabidl1}                                                                 
\end{table} 
%--------------- Table IDL 1

\section{Results}

In this Section we will use some results of our study to
try and answer the following questions on the nature of
soft X-ray emission in obscured AGN:

\begin{itemize}

\item is AGN photoionization the dominant physical process?

\item does resonant scattering play an important role?

\item do efficient X-ray line diagnostics exist, which could allow
us to discriminate on a statistical basis between AGN- and starburst-powered
spectra?

\end{itemize}

\subsection{RRC}

\begin{figure}
\hbox{
\hspace{-1.2cm}
\epsfig{file=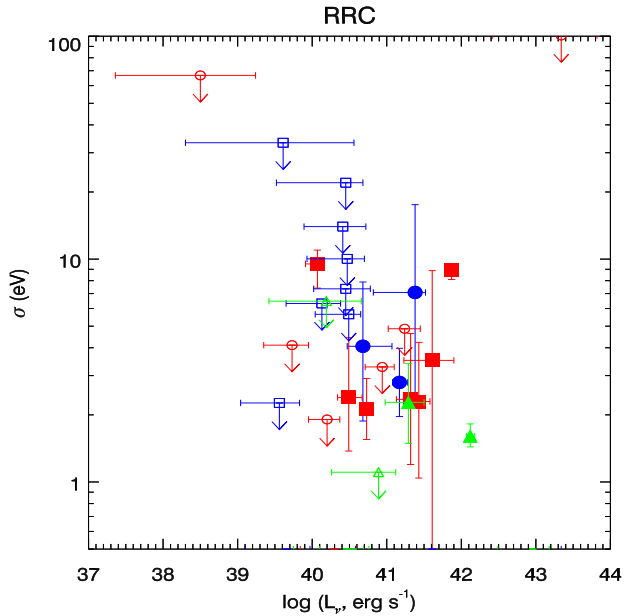,height=85mm,width=95mm}
}
\caption{Luminosity versus
intrinsic width for the RRC features detected in our
sample. {\it Filled}
data points correspond to a 
measurement of the RRC width;
{\it empty} data points correspond
to RRC width upper limits. {\it Circles}:
O{\sc viii}; {\it squares}: O{\sc vii};
{\it triangles}: C{\sc v}.
Data points corresponding to upper
limits on both quantities are not shown
for clarity.}
\label{fig2}
\end{figure}

The three most intense
RRC transitions in the RGS energy bandpass are: O{\sc viii}
at 14.228\AA, O{\sc vii} at 16.771\AA\ and
C{\sc v} at 31.622\AA. At least one of these
features is detected in 36\% (25/69) of the
objects of our sample.
The few measurements of the RRC width,
which represents a direct estimate of
the temperature of the plasma \cite{liedahl96}, cluster
in the range 1--10~eV. There is
an obvious selection effect, favoring the
detection of narrow features. Nonetheless,
still in 29\% (26\%) of the objects in our sample
the RRC width is constrained to be lower than
50 (10)~eV (see Fig.~\ref{fig2}).
%---------------------------------------
%---------------------------------------
In 10 further objects we detect
upper limits on the RRC luminosity in the range
where measurements are found: in 7 (10)
of them
the luminosity is constrained to be
$\le 10^{41.1}$~erg~s$^{-1}$ ($\le 10^{44.2}$~erg~s$^{-1}$).
Finally,
in 34/69 (49\%) of the
objects the quality of the data is too poor
to allow a significant detection of any of the RRC
features,
and the upper limits on their luminosities are
inconclusive.

\begin{figure}
\hbox{
\hspace{-1.2cm}
\epsfig{file=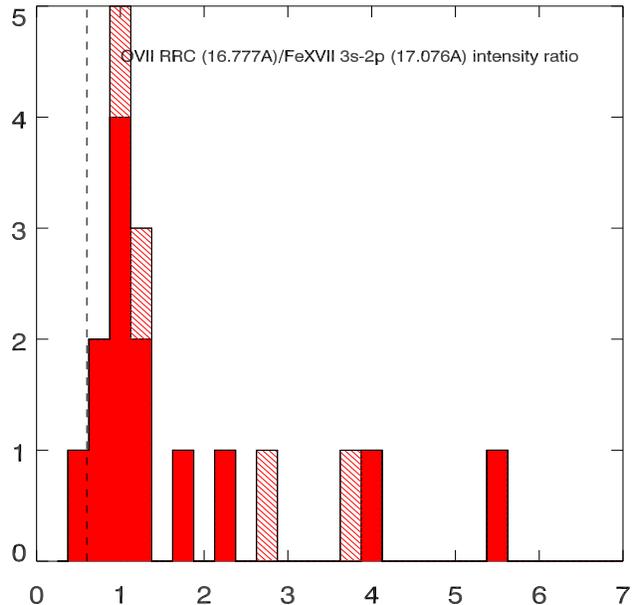,height=85mm,width=95mm}
}
\caption{Observed intensity ratio between the
16.777\AA\ and the 17.076\AA\ features in
{\it CIELO-AGN}. {\it Shaded cells} indicate
lower limits. The {\it dashed line}
represents the maximum expected ratio, if
both features are due to the Fe{\sc xvii}
triplet. In 15 out of 16 cases, the observed ratio
exceeds this limit, indicating that O{\sc vii}
must be the dominant transition at 16.777\AA.}
\label{fig7}
\end{figure}

In principle, measurements of the O{\sc vii} RRC
could be contaminated by the 3F component of
the Fe{\sc xvii} triplet at 16.780\AA\
\cite{brown98}. However, the intensity ratio between
the O{\sc vii} RRC and the 3G/M2 components
of the same triplet (which cannot be resolved
by the RGS) exceeds 0.6, the
largest expected value if the former feature
were entirely due to the 3F component
\cite{phillips97,mauche01,beiersdorfer02},
in 15 out of 16 cases (by more than a factor of 2
in 13 out of 16; cf Fig.~\ref{fig7}). We
%---------------------------------------
%---------------------------------------
are therefore confident that most of our measurements of
the the emission line at 16.777\AA\
represent {\it bona fide} detections of the
O{\sc vii} RRC. The total number of sources
for which at least one RRC feature is
detected is reduced by 1 if the effect of this
potential mis-identification is taken into account.

\subsection{Higher order series}

\begin{figure}
\hbox{
\hspace{-1.2cm}
\epsfig{file=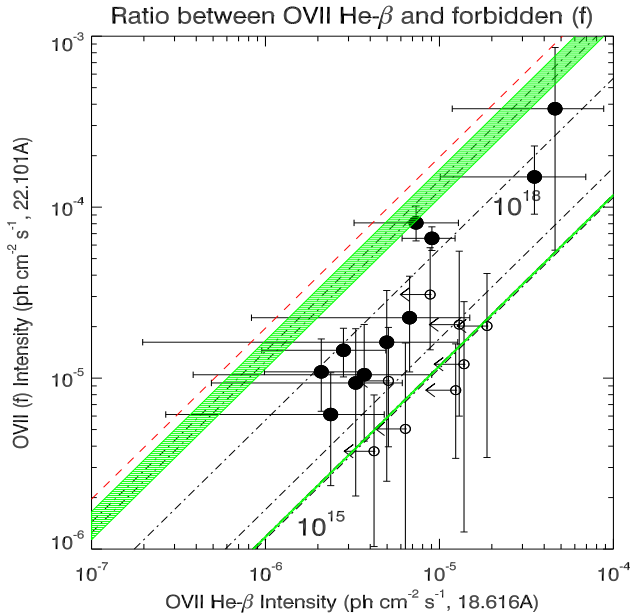,height=85mm,width=95mm}
}
\caption{Intensity of O{\sc vii} He-$\beta$
line against
the intensity of the $f$ component
of the He-$\alpha$ triplet (only
data points corresponding to a detection of the
latter are shown; data points correspond to
upper limits on the intensity of the former
are shown as {\it empty symbols}). The
{\it dashed-dotted lines} represent the prediction
of the {\tt photoion} code for O{\sc vii}
column densities
increasing from $N_{OVII} = 10^{15}$ to $10^{20}$~cm$^{-2}$
in steps of one decade, assuming $kT = 5$~eV and
$v_{turb} = 200$~km~s$^{-1}$.
The {\it long-dashed line} represents the
predictions for pure photoionization.
The {\it shaded areas}
represent the locii of the {\tt photoion}
predictions, when the turbulent velocity
varies in the range 0--500~km~s$^{-1}$
at constant temperature and
for the extreme values of the column density
interval.
The areas representing a variation of the temperature
in the range 1--20~keV at constant velocity and column density
are comparatively smaller, and therefore not shown.}
\label{fig3}
\end{figure}

In addition to $n=2p$$\rightarrow$$1s$ transitions for
H- and He-like atoms, {\it CIELO-AGN} contains a
fair number of detections
of discrete higher order resonance
transitions ($np$$\rightarrow$1$s$, $n > 2$). These
transitions are selectively enhanced by
photoexcitation. Since the forbidden $f$ transition
in He-like triplets is unaffected by photoexcitation,
the intensity ratio between
higher order series and $f$ transition
intensities provides a potentially powerful
diagnostic of the importance of resonant
scattering in radiation ionized spectra
\cite{kinkhabwala02}. An example of the
application of this diagnostic test to {\it CIELO-AGN}
is shown in Fig.~\ref{fig3}, where we display
the intensity of the O{\sc vii} He-$\beta$
against the $f$
component of the O{\sc vii} He-${\alpha}$ triplet. 
%---------------------------------------
%---------------------------------------
We compare the experimental results with
the predictions of pure photoionization, and with
models where radiative decay from photoexcitation
and recombination from photoionization are
self-consistently calculated (model {\tt photoion};
Kinkhabwala et al. 2002). We have produced a grid
of models for different values of
O{\sc vii}
column densities ($N_{OVII} \in$[10$^{15}$,10$^{20}~$~cm$^{-2}$]),
turbulence velocities ($v_{turb} \in$[0,500~km~s$^{-1}$]),
and temperatures ($kT \in$[1,20~keV]).
The weighted mean of the He-$\beta$ versus
$f$ intensity ratio is $0.25 \pm 0.03$,
larger then expected for pure
photoionization.
It corresponds to
O{\sc vii}
column densities $N_{OVII} \simeq$10$^{17-18}$~cm$^{-2}$
(the
dependence on the other parameters is small). 
The measurements on the He-$\beta$ O{\sc vii} transition
at 18.6270\AA\ could be contaminated by the nearby
N{\sc vii} RRC
at 18.5872\AA\footnote{For instance, the
O{\sc vii} He-${\beta}$ width measured in
in NGC~1068, $\sigma = 73 \pm ^5_{13}$~eV (cf. Tab.~\ref{tab2}),
the largest measured in this spectrum,
is most likely contaminated by the N{\sc vii} RRC}.
In order to minimize this contamination,
we fit the X-ray spectra around the O{\sc vii}
He-$\beta$ feature in a range, which in principle does not
include the contaminating feature. Moreover, we have
verified that similar enhancements of higher-order lines to
the Lyman-$\alpha$ of H-like
Carbon and Oxygen are observed as well
(cf. Tab.~\ref{tab1}).
%------------ Table 1
\begin{table}
\caption{Higher-order intensity ratios
for selected transitions in {\it CIELO-AGN}.
Each ratio is calculated as the weighted mean of the
individual ratios on the
$N$ sources, where a measurement of both
transitions is available, and using the
statistical uncertainties on the ratio as weights. Expected
values for pure photoionization (PIE) and
collisional ionization (CIE) are extracted
from Tab.~3 in Kinkhabwala et al. (2002)}
\begin{tabular}{lcccc} \hline \hline
Ion & Ratio & $N$ & PIE & CIE \\ \hline
C{\sc vi} Ly-$\beta$/Ly-$\alpha$ & $0.51 \pm 0.12$ & 9 & 0.14 & 0.09 \\
O{\sc vii} He-$\beta$/$f$ & $0.25 \pm 0.03$ & 12 & 0.05 & ... \\
O{\sc vii} He-$\gamma$/$f$ & $0.20 \pm 0.09$$^a$ & 9 & 0.017 & ... \\
O {\sc viii} Ly-$\beta$/Ly-$\alpha$ & $0.35 \pm 0.05$ & 13 & 0.14 & 0.10 \\ \hline \hline
\end{tabular}

\noindent
$^a$this measurement is likely to be affected by contamination
of Fe{\sc xviii} 3p$^2$-2p$^2$ at $\lambda = 17.8472$~\AA

\label{tab1}
\end{table}
%------------ Table 1
The dependence on the temperature, column density and
velocity distributions
in H-like species is such that no meaningful
constraints on column
density can be drawn. However, in
8 of the individual brightest
sources in our sample\footnote{the Circinus Galaxy, IRAS~13197-1627,
Mrk~3, NGC1068,
NGC1365, NGC2110, NGC2992, NGC4945}
the ratio
between the O{\sc viii} Ly-${\beta}$ and Ly-${\alpha}$ intensities
is large enough to be formally inconsistent with collisional
ionization, and in 4 of them\footnote{NGC1068,
NGC1365, NGC2110, NGC2992} with pure photoionization as well.

\subsection{AGN and starburst soft X-ray spectra}

\begin{figure}
\hbox{
\hspace{-0.6cm}
\epsfig{file=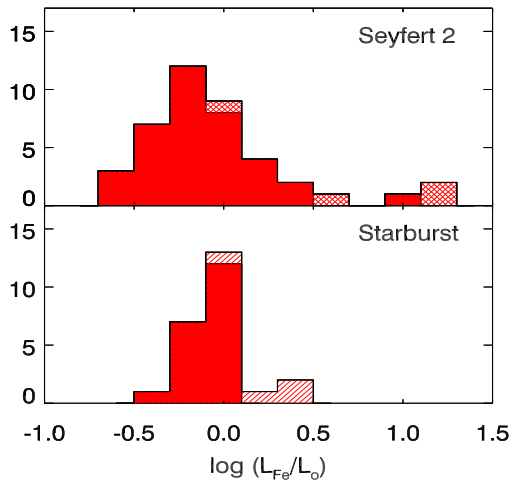,height=75mm,width=85mm}
}
\vspace{-0.5cm}
\caption{Distribution function for
the logarithm of the ratio of the integrated luminosities in
the iron and oxygen lines for our Seyfert~2 ({\it
lower panel}) and in the control sample of starburst
galaxies ({\it lower panel}). {\it Diagonally
shaded cells} indicate upper limits,
{\it checked cells} indicate lower limits. The medians
of the two distributions are very similar:
$-0.10$ and $-0.08$, respectively.}
\label{fig5}
\end{figure}

We have analyzed the RGS spectra of a sample of 27
StarBurst (SB) galaxies extracted from the Wu et al.
(2002) sample (cf. Tab.~\ref{tab3})
%---------------- Table 3
\begin{table}
\begin{tiny}
\begin{center}
\begin{tabular}{lcc} \hline \hline
Source & $z$   & Exposure time \\
       &       & (ks)          \\
ARP244               & 0.006 &  75.0 \\
ARP270               & 0.005 &  36.6 \\
HolmbergII           & 0.000 &  30.6 \\
IZw18                & 0.003 &  52.3 \\
M82                  & 0.001 & 141.6 \\
M83                  & 0.002 &  30.5 \\
MCG-5-31-7           & 0.010 &  19.0 \\
NGC1313              & 0.002 & 241.4 \\
NGC1482              & 0.006 &  17.9 \\
NGC1511              & 0.004 &  44.0 \\
NGC1705              & 0.002 &  58.5 \\
NGC2146              & 0.003 &  35.9 \\
NGC2403              & 0.000 &  43.5 \\
NGC253               & 0.001 & 192.3 \\
NGC2798              & 0.006 &  13.8 \\
NGC3256              & 0.009 &  16.7 \\
NGC3310              & 0.003 &  19.2 \\
NGC3690              & 0.010 &  21.5 \\
NGC4214              & 0.001 &  19.5 \\
NGC4449              & 0.004 &  28.1 \\
NGC474               & 0.008 &  25.4 \\
NGC5073              & 0.009 &  51.8 \\
NGC520               & 0.008 &  12.6 \\
NGC5253              & 0.001 &  45.8 \\
NGC660               & 0.003 &  11.9 \\
NGC7552              & 0.005 &  24.0 \\
NGC7714              & 0.009 &  41.3 \\
\hline \hline
\end{tabular}
\end{center}
\end{tiny}
\caption{Log of the XMM-Newton observations
of the starburst galaxies in the control sample}
\label{tab3}
\end{table}
%---------------- Table 3
and observed by XMM-Newton, to identify
diagnostic criteria, which may allow us to statistically
discriminate between SB- and AGN-dominated soft
X-ray spectra.
This sample has been analyzed in the same way as the
Seyfert~2 sample.
Some of the criteria quoted in the literature
provide inconclusive results. For instance,
the distribution functions of the
luminosity ratio between L shell iron
and integrated oxygen He- and H-like lines are
indistinguishable (Fig.~\ref{fig5}).
Kallman et al. (1996) had already pointed out that
L-shell iron transitions can give an important
contribution to the overall luminosity budget in
photoionized nebulae.
%---------------------------------------
%---------------------------------------
The standard $G$  ratio
($G \equiv (f+i)/r$;
Gabriel \& Jordan 1969; Porquet et al. 2000),
is also ambiguous.
Previous studies have shown (c.f. Sect.~1) that the enhancement of resonant lines
by resonant scattering mimics the behavior of a collisionally ionized plasma.
In the collisionally ionized case, the fluxes of all the triplet lines are enhanced,
with the greatest increase in the resonance line.
Resonant scattering boosts all of the resonant lines with respect to the
forbidden and intercombination lines in the triplet.
The value of G is therefore decreased by both collisional excitation
and resonant scattering.
Finally, the fraction of SB spectra
where narrow RRC features are detected is comparable
to that observed in the Seyfert~2 sample,
indicating that photoionization plays an important
role in the ionization balance of starburst galaxies as well.

\begin{figure}
\hbox{
\hspace{-1.2cm}
\epsfig{file=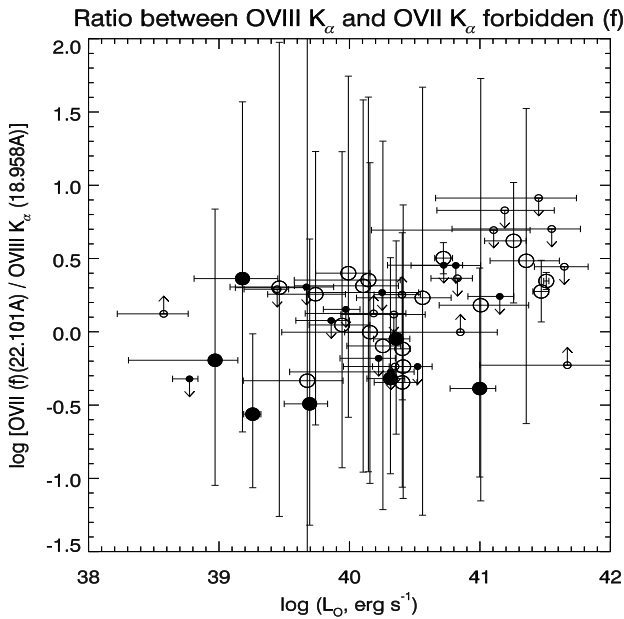,height=85mm,width=95mm}
}
\caption{Intensity of the $f$
component of the O{\sc vii} triplet
(normalized to the O{\sc viii} Ly-${\alpha}$
intensity)
against the total luminosity in Oxygen lines.
{\it Empty circles} represents the obscured
AGN in {\it CIELO-AGN}, {\it filled circles} the
control sample of 27 starburst galaxies.
Symbols representing
censored data are plotted with a smaller size for
the sake of clarity.
}
\label{fig4}
\end{figure}

In Fig.~\ref{fig4} we compare the distribution of
the O{\sc vii} $f$ transition intensity,
normalized against the intensity of the O{\sc viii} Ly-${\alpha}$,
as a function of the total luminosity in oxygen lines (integrated on
all He- and H-like transitions).
%---------------------------------------
%---------------------------------------
AGN are generally characterized by
larger intensity ratios, $\eta$:
$\eta_{AGN} = 1.38 \pm^{0.13}_{0.12}$,
$\eta_{SB} = 0.57 \pm^{0.07}_{0.09}$
(5.8$\sigma$ difference; the upper limits have been taken into
account with a 
``bootstrap'' method for censored data as in
Schmitt 1985). The same figure also shows that the
average line luminosity in AGN is
larger than in starbursts. The median
Oxygen lines luminosity is 
$\sim$10$^{41}$~erg~s$^{-1}$ in the former
and $\sim$10$^{40}$~erg~s$^{-1}$ in the latter
(using strict detections only).
Still, the overlap
between the line diagnostic distributions is 
significant, and prevents strong statements 
on individual sources. This most likely
reflects the intrinsic ``composite'' nature of several
obscured AGN \cite{cidfernandes01}.

\section{Conclusions}

In this paper we present results from a systematic study
of high-resolution soft X-ray spectra for 69 obscured AGN
observed with the XMM-Newton RGS. Our main conclusions can
be summarized as follows: 

\begin{itemize}

\item[-]
\underline{Radiative Recombination Continua:} we detect
RRC features in about 40\% of the sample sources, and in
33\% (27\%) they allow us to constraint the
temperature of the plasma $\le 50$~eV ($\le 10$)~eV.
This indicates that in a large fraction of
objects in our sample photoionization dominates the
ionization balance.
Still, these percentages represent probably lower limits to the
true fraction of photoionized sources in our
sample: upper limits on narrow RRC features in
almost 50\% of the objects in our sample are
still inconclusive due to the lack of
statistics

\item[-]
\underline{Resonant scattering:} higher order
transitions are enhanced with respect to the expectation
of pure photoionization, and inconsistent
with the expectation of collisional ionization as well. This indicates that
resonant scattering plays an important role
in the ionization/excitation balance. The observed O{\sc vii}
He-$\beta$ versus $f$ intensity ratios are consistent with
O{\sc vii}
column densities in the range $N_{OVII} \sim$10$^{17-18}$~cm$^{-2}$.
Interestingly enough, these values are
in good agreement with the column densities measured in
``warm absorbers'' observed along the lines of sight to
unobscured AGN \cite{blustin03,steenbrugge05,blustin05}.

\item[-]
\underline{Starburst contribution:} the comparison between
the spectra of obscured AGN in our sample and a control
sample of nearby starburst galaxies suggests two empirical
criteria to discriminate {\it on a statistical basis} between AGN-
and
starburst-powered spectra: a) total Oxygen
line X-ray luminosity $\approxgt 10^{40}$~erg~s$^{-1}$; b)
ratio between the O{\sc vii} triplet
$f$ component and the O{\sc viii}
Ly-${\alpha}$~$\approxgt$1.

\end{itemize}

The currently available X-ray instrumentation allowed
us to explore the nature of the soft X-ray emission
in obscured AGN as weak as $\sim$0.03~mCrab.
From this study, we have gained confidence that
conclusions on the properties of the gas in
the circumnuclear region of AGN extracted
from the detailed study
of high-quality spectra of the brightest objects
can be extended to the whole population of nearby obscured
AGN.
This is the main message we would like to convey with this
{\it paper}, which opens interesting perspectives for future
enlargements of this study once deeper exposures or more
sensitive instrumentation are available.

\section*{Appendix}

The tables in this
Section list centroid energies and fluxes for the 
bound-bound transitions individually discussed in
this paper: C{\sc vi} Ly-$\alpha$,
C{\sc vi} Ly-$\beta$,
O{\sc vii} He-$\alpha$,
O{\sc vii} He-$\beta$,
O{\sc viii} Ly-$\alpha$, and O{\sc viii} Ly-$\beta$
(Tab.~\ref{tabidl2} to \ref{tabidl3}).
%------------------------- Table 2 IDL
\begin{table*}                                                                  
\begin{tiny}                                                                    
\begin{tabular}{lcccccc} \hline \hline                                          
Source & C{\sc vi} Ly-$\alpha$ & C{\sc vi} Ly-$\beta$ & O{\sc vii} He-$\alpha$~(r) & O{\sc vii} He-$\beta$ & O{\sc viii} Ly-$\alpha$ & O{\sc viii} Ly-$\beta$ \\                    
 & (\AA, 33.737) & (\AA, 28.459) & (\AA, 21.602) & (\AA, 18.627) & (\AA, 18.967) & (\AA, 16.006) \\ \hline                                                                          
Circinus~Galaxy                       & ...                            & ...                            & $21.689\pm^{0.754}_{0.560}$    & $18.616\pm^{0.052}_{0.054}$    & $18.941\pm^{0.040}_{0.046}$    & $15.999\pm^{0.048}_{0.047}$    \\
ESO509-G66                     & ...                            & ...                            & ...                            & ...                            & ...                            & ...                            \\
IC2560                         & ...                            & ...                            & $21.598\pm^{0.543}_{0.518}$    & $18.640\pm^{0.054}_{0.047}$    & ...                            & $16.075\pm^{0.087}_{0.060}$    \\
IC4395                         & ...                            & ...                            & ...                            & ...                            & ...                            & ...                            \\
IC4995                         & $33.430\pm^{0.086}_{0.172}$    & ...                            & ...                            & ...                            & ...                            & ...                            \\
IIIZW035                       & ...                            & ...                            & ...                            & ...                            & ...                            & ...                            \\
IRAS01475-0740                 & ...                            & ...                            & ...                            & ...                            & ...                            & ...                            \\
IRAS08572+3915                 & ...                            & ...                            & ...                            & ...                            & ...                            & ...                            \\
IRAS09104+4109                 & ...                            & ...                            & ...                            & ...                            & ...                            & ...                            \\
IRAS10214+4724                 & ...                            & ...                            & ...                            & ...                            & ...                            & ...                            \\
IRAS13197-1627                 & ...                            & ...                            & ...                            & $18.700\pm^{0.177}_{0.175}$    & $18.931\pm^{0.050}_{0.050}$    & $16.015\pm^{0.049}_{0.055}$    \\
IRAS15480-0344                 & ...                            & ...                            & ...                            & ...                            & ...                            & ...                            \\
MCG-5-23-16                    & ...                            & ...                            & $21.597\pm^{0.557}_{0.527}$    & $18.664\pm^{0.074}_{0.081}$    & ...                            & ...                            \\
MRK3                           & ...                            & ...                            & ...                            & $18.615\pm^{0.038}_{0.038}$    & ...                            & $16.022\pm^{0.061}_{0.048}$    \\
MRK6                           & ...                            & ...                            & ...                            & ...                            & ...                            & ...                            \\
MRK331                         & ...                            & ...                            & ...                            & $18.367\pm^{0.392}_{0.377}$    & ...                            & ...                            \\
MRK348                         & $33.415\pm^{0.074}_{0.106}$    & ...                            & ...                            & ...                            & $19.013\pm^{0.115}_{0.125}$    & ...                            \\
MRK612                         & ...                            & ...                            & ...                            & ...                            & ...                            & ...                            \\
MRK744                         & ...                            & ...                            & ...                            & ...                            & ...                            & ...                            \\
MRK993                         & ...                            & ...                            & ...                            & ...                            & ...                            & ...                            \\
MRK1152                        & ...                            & ...                            & ...                            & ...                            & ...                            & ...                            \\
NGC1068                        & ...                            & ...                            & $21.576\pm^{0.029}_{0.029}$    & $18.685\pm^{0.044}_{0.039}$    & $18.962\pm^{0.029}_{0.029}$    & $16.006\pm^{0.033}_{0.032}$    \\
NGC1365                        & $33.767\pm^{0.052}_{0.050}$    & $28.491\pm^{0.053}_{0.052}$    & $21.600\pm^{0.029}_{0.029}$    & $18.634\pm^{0.052}_{0.054}$    & $18.966\pm^{0.030}_{0.030}$    & $16.030\pm^{0.047}_{0.047}$    \\
NGC1386                        & $33.776\pm^{0.052}_{0.064}$    & ...                            & $21.592\pm^{0.535}_{0.522}$    & ...                            & $18.933\pm^{0.065}_{0.053}$    & ...                            \\
NGC1410                        & ...                            & ...                            & ...                            & ...                            & ...                            & ...                            \\
NGC1614                        & ...                            & ...                            & ...                            & $18.598\pm^{0.183}_{0.181}$    & ...                            & ...                            \\
NGC2110                        & $33.397\pm^{0.065}_{0.069}$    & ...                            & ...                            & $18.663\pm^{0.071}_{0.074}$    & $18.994\pm^{0.047}_{0.044}$    & $16.181\pm^{0.051}_{0.045}$    \\
NGC2273                        & ...                            & ...                            & ...                            & ...                            & ...                            & ...                            \\
NGC2623                        & ...                            & ...                            & ...                            & ...                            & ...                            & ...                            \\
NGC2992                        & ...                            & $28.076\pm^{0.059}_{0.076}$    & $21.598\pm^{0.051}_{0.052}$    & ...                            & $18.941\pm^{0.053}_{0.072}$    & $15.802\pm^{0.054}_{0.071}$    \\
NGC34                          & ...                            & ...                            & ...                            & ...                            & ...                            & ...                            \\
NGC3982                        & ...                            & ...                            & ...                            & ...                            & $18.813\pm^{0.052}_{0.068}$    & ...                            \\
NGC4138                        & $33.988\pm^{0.063}_{0.063}$    & $28.400\pm^{0.298}_{0.107}$    & $21.563\pm^{0.658}_{0.607}$    & ...                            & $19.245\pm^{0.072}_{0.072}$    & ...                            \\
NGC4151                        & $33.740\pm^{0.033}_{0.035}$    & $28.439\pm^{0.095}_{0.049}$    & ...                            & ...                            & ...                            & ...                            \\
NGC4168                        & ...                            & ...                            & ...                            & $18.361\pm^{0.053}_{0.053}$    & ...                            & $15.779\pm^{0.062}_{0.062}$    \\
NGC424                         & $33.733\pm^{0.118}_{0.097}$    & ...                            & $21.612\pm^{0.552}_{0.524}$    & ...                            & $18.923\pm^{0.065}_{0.049}$    & ...                            \\
NGC4258                        & $33.617\pm^{0.117}_{0.148}$    & $28.506\pm^{0.047}_{0.047}$    & ...                            & ...                            & $19.044\pm^{0.061}_{0.081}$    & $15.961\pm^{0.187}_{0.100}$    \\
NGC4303                        & ...                            & ...                            & $21.636\pm^{0.050}_{0.051}$    & ...                            & $18.971\pm^{0.053}_{0.049}$    & ...                            \\
NGC4395                        & $33.895\pm^{0.066}_{0.050}$    & ...                            & $21.589\pm^{0.545}_{0.518}$    & $18.740\pm^{0.092}_{0.090}$    & ...                            & $15.819\pm^{0.052}_{0.050}$    \\
NGC4472                        & ...                            & ...                            & ...                            & ...                            & ...                            & $16.078\pm^{0.053}_{0.047}$    \\
NGC4477                        & $33.747\pm^{0.068}_{0.080}$    & $28.152\pm^{0.086}_{0.086}$    & ...                            & ...                            & $18.992\pm^{0.072}_{0.073}$    & ...                            \\
NGC449                         & ...                            & ...                            & $21.602\pm^{0.561}_{0.518}$    & ...                            & ...                            & ...                            \\
NGC4507                        & $34.025\pm^{0.830}_{0.793}$    & ...                            & $21.601\pm^{0.029}_{0.029}$    & $18.651\pm^{0.062}_{0.046}$    & $18.958\pm^{0.044}_{0.033}$    & $16.018\pm^{0.109}_{0.073}$    \\
NGC4565                        & ...                            & ...                            & ...                            & $18.719\pm^{0.058}_{0.046}$    & ...                            & $15.857\pm^{0.179}_{0.176}$    \\
NGC4639                        & ...                            & $28.549\pm^{0.084}_{0.065}$    & ...                            & ...                            & ...                            & ...                            \\
NGC4725                        & $33.731\pm^{0.231}_{0.201}$    & $28.236\pm^{0.052}_{0.103}$    & $21.589\pm^{0.554}_{0.547}$    & ...                            & $18.974\pm^{0.038}_{0.053}$    & ...                            \\
NGC4945                        & ...                            & ...                            & ...                            & ...                            & $18.986\pm^{0.080}_{0.048}$    & $16.104\pm^{0.074}_{0.057}$    \\
NGC4968                        & ...                            & ...                            & ...                            & ...                            & ...                            & $16.074\pm^{0.052}_{0.047}$    \\
NGC5033                        & ...                            & $28.657\pm^{0.053}_{0.045}$    & $21.586\pm^{0.537}_{0.523}$    & ...                            & $19.025\pm^{0.048}_{0.050}$    & $16.178\pm^{0.109}_{0.108}$    \\
NGC5252                        & $33.370\pm^{0.112}_{0.098}$    & ...                            & ...                            & ...                            & $18.896\pm^{0.153}_{0.151}$    & ...                            \\
NGC526A                        & ...                            & ...                            & ...                            & ...                            & $18.928\pm^{0.119}_{0.118}$    & ...                            \\
NGC5273                        & ...                            & $28.668\pm^{0.092}_{0.068}$    & $21.609\pm^{0.530}_{0.516}$    & $18.556\pm^{0.129}_{0.069}$    & ...                            & ...                            \\
NGC5506                        & $33.819\pm^{0.039}_{0.060}$    & $28.375\pm^{0.041}_{0.048}$    & $21.573\pm^{0.059}_{0.058}$    & $18.632\pm^{0.121}_{0.082}$    & ...                            & $16.001\pm^{0.110}_{0.066}$    \\
NGC5643                        & $34.018\pm^{0.123}_{0.101}$    & ...                            & ...                            & ...                            & $18.976\pm^{0.053}_{0.043}$    & ...                            \\
NGC591                         & ...                            & ...                            & ...                            & ...                            & $18.974\pm^{0.125}_{0.052}$    & ...                            \\
NGC6552                        & ...                            & ...                            & ...                            & ...                            & ...                            & ...                            \\
NGC7172                        & ...                            & ...                            & $21.566\pm^{0.042}_{0.042}$    & ...                            & $18.977\pm^{0.076}_{0.076}$    & ...                            \\
NGC7212                        & ...                            & ...                            & ...                            & ...                            & ...                            & ...                            \\
NGC7314                        & ...                            & ...                            & $21.602\pm^{0.047}_{0.048}$    & ...                            & $18.868\pm^{0.080}_{0.093}$    & ...                            \\
NGC7479                        & ...                            & ...                            & ...                            & ...                            & $18.948\pm^{0.085}_{0.238}$    & ...                            \\
NGC7582                        & $33.676\pm^{0.059}_{0.064}$    & ...                            & $21.603\pm^{0.049}_{0.048}$    & $18.624\pm^{0.123}_{0.053}$    & $18.960\pm^{0.043}_{0.034}$    & ...                            \\
NGC7674                        & ...                            & ...                            & ...                            & ...                            & ...                            & ...                            \\
UGC1214                        & $33.545\pm^{0.057}_{0.069}$    & $28.349\pm^{0.102}_{0.081}$    & ...                            & ...                            & $18.956\pm^{0.081}_{0.053}$    & $16.065\pm^{0.141}_{0.139}$    \\
UGC2456                        & $33.595\pm^{0.066}_{0.074}$    & ...                            & ...                            & $18.634\pm^{0.073}_{0.079}$    & ...                            & $15.836\pm^{0.054}_{0.044}$    \\
UGC2608                        & ...                            & ...                            & ...                            & ...                            & ...                            & ...                            \\
UGC4203                        & $33.797\pm^{0.122}_{0.072}$    & ...                            & ...                            & ...                            & ...                            & $16.022\pm^{0.103}_{0.070}$    \\
UGC6527                        & ...                            & ...                            & ...                            & ...                            & ...                            & ...                            \\
UGC8621                        & ...                            & ...                            & ...                            & ...                            & ...                            & ...                            \\
UM625                          & ...                            & ...                            & ...                            & ...                            & ...                            & ...                            \\
\hline \hline                                                                   
\end{tabular}                                                                   
\end{tiny}                                                                      
\caption{Best-fit wavelengths for the transitions discussed in this             
{\it paper}. {\it Dots} indicates either lines not detected, or                 
lines for which no constraints can be obtained (the luminosities in Tab.~\ref{tabidl3}              
are calculated in this case assuming the                                        
laboratory energies). In the table header, the laboratory wavelengths are in       
brackets.}                                                                      
\label{tabidl2}                                                                 
\end{table*}         
%------------------------- Table 2 IDL
%------------------------- Table 3 IDL
\begin{table*}                                                                  
\begin{tiny}                                                                    
\begin{tabular}{lcccccc} \hline \hline                                          
Source & C{\sc vi} Ly-$\alpha$ & C{\sc vi} Ly-$\beta$ & O{\sc vii} He-$\alpha$ & O{\sc vii} He-$\beta$ & O{\sc viii} Ly-$\alpha$ & O{\sc viii} Ly-$\beta$ \\                        
 & $10^{-5}$~ph~cm$^{-2}$~s$^{-1}$ & $10^{-5}$~ph~cm$^{-2}$~s$^{-1}$ & $10^{-5}$~ph~cm$^{-2}$~s$^{-1}$ & $10^{-5}$~ph~cm$^{-2}$~s$^{-1}$ & $10^{-5}$~ph~cm$^{-2}$~s$^{-1}$ & $10^{-5}$~ph~cm$^{-2}$~s$^{-1}$ \\ \hline                                    
Circinus~Galaxy             & ...        & ...                        & $37.6\pm^{48.2}_{32.0}$    & $4.6\pm^{4.1}_{3.4}$       & $14.9\pm^{7.3}_{6.0}$      & $3.6\pm^{1.9}_{1.6}$       \\
ESO509-G66                 & ...                                      & ...                        & ...                        & ...                        & ...                        & ...                        \\
IC2560                     & ...                                      & ...                        & $1.0\pm^{0.6}_{0.4}$       & $0.2\pm^{0.1}_{0.1}$       & ...                        & $0.1\pm^{0.1}_{0.0}$       \\
IC4395                     & ...                                      & ...                        & ...                        & ...                        & ...                        & ...                        \\
IC4995                     & $1.6\pm^{2.4}_{1.4}$                     & ...                        & ...                        & ...                        & ...                        & ...                        \\
IIIZW035                   & ...                                      & ...                        & ...                        & ...                        & ...                        & ...                        \\
IRAS01475-0740             & ...                                      & ...                        & ...                        & ...                        & ...                        & ...                        \\
IRAS08572+3915             & ...                                      & ...                        & ...                        & ...                        & ...                        & ...                        \\
IRAS09104+4109             & ...                                      & ...                        & ...                        & ...                        & ...                        & ...                        \\
IRAS10214+4724            & ...                                      & ...                        & ...                        & ...                        & ...                        & ...                        \\
IRAS13197-1627             & ...                                      & ...                        & $15.0\pm^{7.7}_{5.9}$      & $3.5\pm^{3.4}_{2.5}$       & $1.0\pm^{0.5}_{0.4}$       & $0.4\pm^{0.3}_{0.2}$       \\
IRAS15480-0344             & ...                                      & ...                        & ...                        & ...                        & ...                        & ...                        \\
MCG-5-23-16                & ...                                      & ...                        & $1.6\pm^{1.6}_{1.3}$       & $0.4\pm^{0.7}_{0.4}$       & ...                        & ...                        \\
MRK3                 & ...                                      & ...                        & $6.5\pm^{1.0}_{0.9}$       & $0.9\pm^{0.3}_{0.2}$       & $3.4\pm^{0.4}_{0.4}$       & $0.7\pm^{0.2}_{0.2}$       \\
MRK6                 & ...                                      & ...                        & ...                        & ...                        & ...                        & ...                        \\
MRK331                     & ...                                      & ...                        & ...                        & $0.4\pm^{0.6}_{0.3}$       & ...                        & ...                        \\
MRK348                     & $2.5\pm^{2.3}_{1.8}$                     & ...                        & ...                        & ...                        & $1.2\pm^{1.1}_{1.0}$       & ...                        \\
MRK612                     & ...                                      & ...                        & ...                        & ...                        & ...                        & ...                        \\
MRK744                     & ...                                & ...                        & ...                        & ...                        & ...                        & ...                        \\
MRK993                     & ...                                      & ...                        & ...                        & ...                        & ...                        & ...                        \\
MRK1152                    & ...                                      & ...                        & ...                        & ...                        & ...                        & ... \\
NGC1068                    & $98.6\pm^{4.6}_{4.5}$                    & $10.9\pm^{1.5}_{1.5}$      & $124.9\pm^{4.6}_{3.6}$     & $44.3\pm^{3.3}_{3.8}$      & $53.2\pm^{2.1}_{2.1}$      & $14.9\pm^{1.2}_{1.1}$      \\
NGC1365                    & $1.1\pm^{0.5}_{0.4}$                     & $0.4\pm^{0.2}_{0.1}$       & $1.4\pm^{0.5}_{0.4}$       & $0.2\pm^{0.2}_{0.1}$       & $1.9\pm^{0.3}_{0.3}$       & $0.5\pm^{0.2}_{0.1}$       \\
NGC1386                    & $0.9\pm^{1.1}_{0.7}$                     & ...                        & $1.8\pm^{1.5}_{0.9}$       & ...                        & $1.0\pm^{0.7}_{0.5}$       & ...                        \\
NGC1410                    & ...                                      & ...                        & ...                        & ...                        & ...                        & ...                        \\
NGC1614                    & ...                                      & ...                        & ...                        & $0.4\pm^{0.5}_{0.4}$       & ...                        & ...                        \\
NGC2110                    & $9.6\pm^{10.1}_{7.1}$                    & ...                        & ...                        & $0.6\pm^{0.7}_{0.5}$       & $1.2\pm^{0.8}_{0.6}$       & $1.5\pm^{0.8}_{0.6}$       \\
NGC2273                    & ...                                      & ...                        & ...                        & ...                        & ...                        & ...                        \\
NGC2623                    & ...                                      & ...                        & ...                        & ...                        & ...                        & ...                        \\
NGC2992                    & ...                                      & $1.0\pm^{0.8}_{0.7}$       & $1.2\pm^{1.5}_{1.0}$       & ...                        & $0.7\pm^{0.7}_{0.6}$       & $1.2\pm^{1.0}_{0.8}$       \\
NGC34                      & ...                                      & ...                        & ...                        & ...                        & ...                        & ...                        \\
NGC3982                    & ...                                      & ...                        & $1.4\pm^{1.3}_{1.1}$       & ...                        & $0.7\pm^{0.7}_{0.5}$       & ...                        \\
NGC4138                    & $0.9\pm^{1.2}_{0.6}$                     & $0.4\pm^{0.7}_{0.4}$       & $0.5\pm^{1.1}_{0.4}$       & ...                        & $1.0\pm^{1.7}_{0.9}$       & ...                        \\
NGC4151                    & $17.7\pm^{2.1}_{2.0}$                    & $2.0\pm^{0.8}_{0.7}$       & $49.5\pm^{3.3}_{3.1}$      & $1.3\pm^{0.8}_{0.7}$       & $15.5\pm^{1.3}_{1.2}$      & $2.5\pm^{0.6}_{0.5}$       \\
NGC4168                    & ...                                      & ...                        & ...                        & $0.2\pm^{0.3}_{0.2}$       & ...                        & $0.2\pm^{0.4}_{0.2}$       \\
NGC424                     & $2.0\pm^{2.3}_{1.4}$                     & ...                        & $3.0\pm^{2.7}_{1.6}$       & ...                        & $1.0\pm^{0.9}_{0.5}$       & ...                        \\
NGC4258                    & $0.6\pm^{0.6}_{0.5}$                     & $0.5\pm^{0.4}_{0.3}$       & ...                        & $1.9\pm^{0.6}_{0.5}$       & $0.6\pm^{0.4}_{0.4}$       & $0.3\pm^{0.3}_{0.3}$       \\
NGC4303                    & $41.6\pm^{29.9}_{32.8}$                  & ...                        & $0.5\pm^{0.7}_{0.5}$       & ...                        & $0.7\pm^{0.4}_{0.4}$       & ...                        \\
NGC4395                    & $0.6\pm^{0.4}_{0.3}$                     & ...                        & $0.6\pm^{0.4}_{0.3}$       & $0.2\pm^{0.2}_{0.2}$       & ...                        & $0.3\pm^{0.2}_{0.2}$       \\
NGC4472                    & ...                                      & ...                        & ...                        & ...                        & ...                        & $7.9\pm^{7.0}_{4.9}$       \\
NGC4477                    & $1.6\pm^{1.9}_{1.2}$                     & $0.9\pm^{1.2}_{0.8}$       & ...                        & ...                        & $0.8\pm^{0.9}_{0.6}$       & ...                        \\
NGC449                     & ...                                      & ...                        & $2.0\pm^{3.4}_{1.4}$       & ...                        & ...                        & ...                        \\
NGC4507                    & $67.6\pm^{65.2}_{65.2}$                  & ...                        & $8.0\pm^{2.0}_{1.7}$       & $0.7\pm^{0.5}_{0.4}$       & $1.9\pm^{0.7}_{0.5}$       & $0.2\pm^{0.3}_{0.2}$       \\
NGC4565                    & ...                                      & ...                        & ...                        & $0.2\pm^{0.3}_{0.1}$       & ...                        & $0.3\pm^{0.4}_{0.2}$       \\
NGC4639                    & ...                                      & $0.7\pm^{1.0}_{0.6}$       & ...                        & ...                        & ...                        & ...                        \\
NGC4725                    & $0.7\pm^{1.1}_{0.7}$                     & $0.5\pm^{0.7}_{0.4}$       & $0.9\pm^{1.0}_{0.5}$       & ...                        & $0.8\pm^{0.6}_{0.4}$       & ...                        \\
NGC4945                    & $9.3\pm^{12.5}_{8.0}$                    & ...                        & $1.9\pm^{2.8}_{1.5}$       & ...                        & $0.9\pm^{1.3}_{0.8}$       & $0.4\pm^{0.5}_{0.3}$       \\
NGC4968                    & ...              & ...                        & ...                        & ...                        & ...                        & $1.2\pm^{1.1}_{0.7}$       \\
NGC5033                    & ...                                      & $2.4\pm^{1.4}_{1.3}$       & $2.0\pm^{2.0}_{1.6}$       & ...                        & $2.0\pm^{1.2}_{1.0}$       & $0.9\pm^{1.0}_{0.8}$       \\
NGC5252                    & $0.6\pm^{0.5}_{0.6}$                     & ...                        & ...                        & ...                        & $2.0\pm^{2.3}_{1.8}$       & ...                        \\
NGC526A                    & ...                                      & ...                        & ...                        & ...                        & $0.3\pm^{0.5}_{0.3}$       & ...                        \\
NGC5273                    & ...                                      & $0.8\pm^{0.9}_{0.6}$       & $2.2\pm^{1.6}_{1.1}$       & $0.6\pm^{0.8}_{0.5}$       & ...                        & ...                        \\
NGC5506                    & $2.0\pm^{1.3}_{1.3}$                     & $0.7\pm^{0.6}_{0.3}$       & $0.9\pm^{1.0}_{0.7}$       & $0.3\pm^{0.2}_{0.2}$       & $1.6\pm^{0.7}_{0.3}$       & $0.4\pm^{0.3}_{0.3}$       \\
NGC5643                    & $2.9\pm^{5.1}_{2.8}$                     & ...                        & ...                        & ...                        & $2.6\pm^{1.8}_{1.3}$       & ...                        \\
NGC591                     & ...                                      & ...                        & ...                        & ...                        & $0.6\pm^{0.9}_{0.4}$       & ...                        \\
NGC6552                    & ...                                      & ...                        & ...                        & ...                        & ...                        & ...                        \\
NGC7172                    & ...                                      & ...                        & $0.3\pm^{0.4}_{0.2}$       & ...                        & $0.2\pm^{0.2}_{0.1}$       & ...                        \\
NGC7212                    & ...                                      & ...                        & ...                        & ...                        & ...                        & ...                        \\
NGC7314                    & ...                                      & ...                        & $0.8\pm^{0.7}_{0.5}$       & ...                        & $0.3\pm^{0.4}_{0.3}$       & ...                        \\
NGC7479                    & ...                                      & ...                        & ...                        & ...                        & $0.7\pm^{1.0}_{0.7}$       & ...                        \\
NGC7582                    & $1.8\pm^{1.6}_{1.1}$                     & ...                        & $1.0\pm^{1.0}_{0.6}$       & $0.3\pm^{0.5}_{0.3}$       & $2.3\pm^{0.9}_{0.7}$       & ...                        \\
NGC7674                    & ...                                      & ...                        & ...                        & ...                        & ...                        & ...                        \\
UGC1214                    & $1.7\pm^{1.8}_{1.1}$                     & $1.6\pm^{2.2}_{1.1}$       & ...                        & ...                        & $3.3\pm^{2.1}_{1.5}$       & $0.8\pm^{1.2}_{0.5}$       \\
UGC2456                    & $9.8\pm^{11.0}_{7.6}$                    & ...                        & ...                        & $0.9\pm^{1.1}_{0.7}$       & ...                        & $1.1\pm^{0.9}_{0.7}$       \\
UGC2608                    & ...                                      & ...                        & ...                        & ...                        & ...                        & ...                        \\
UGC4203                    & $12.5\pm^{13.9}_{9.6}$                   & ...                        & ...                        & ...                        & ...                        & $1.7\pm^{2.0}_{1.5}$       \\
UGC6527                    & ...                                      & ...                        & ...                        & ...                        & ...                        & ...                        \\
UGC8621                    & ...                                      & ...                        & ...                        & ...                        & ...                        & ...                        \\
UM625                      & ...                                      & $1.0\pm^{1.1}_{0.8}$       & ...                        & ...                        & ...                        & ...                        \\
\hline \hline                                                                   
\end{tabular}                                                                   
\end{tiny}                                                                      
\caption{Best-fit fluxes for the transitions discussed in this {\it paper}.     
{\it Dots} indicate lines not detected.}                                       
\label{tabidl3}                                                                 
\end{table*}                                                                    
%------------------------- Table 3 IDL
The same
quantities are listed for the C{\sc v}, O{\sc vii},
and O{\sc viii} RRCs, alongside their intrinsic width
(Tab.~\ref{tabidl4} to \ref{tabidl5}).
%------------------------- Table 4 IDL
\begin{table*}                                                                  
\begin{tiny}                                                                    
\begin{tabular}{lcccccc} \hline \hline                                          
Source & \multicolumn{2}{c}{C{\sc v} RRC} &  \multicolumn{2}{c}{O{\sc vii} RRC} &  \multicolumn{2}{c}{O{\sc viii} RRC} \\                                                           
 & $E_c$ & $\sigma$ & $E_c$ & $\sigma$ & $E_c$ & $\sigma$  \\                   
 & (\AA, 31.622) & (eV) & (\AA, 16.771) & (eV) & (\AA, 14.228) & (eV) \\ \hline 
Circinus~Galaxy                       & ...                            & ...                            & $16.775\pm^{0.169}_{0.164}$    & $<33.2$                        & $14.182\pm^{0.058}_{0.040}$    & $<4.1$                         \\
ESO509-G66                     & ...                            & ...                            & ...                            & ...                            & ...                            & ...                            \\
IC2560                         & ...                            & ...                            & $16.681\pm^{0.102}_{0.066}$    & $<10.0$                        & ...                            & ...                            \\
IC4395                         & ...                            & ...                            & ...                            & ...                            & ...                            & ...                            \\
IC4995                         & ...                            & ...                            & ...                            & ...                            & ...                            & ...                            \\
IIIZW035                       & ...                            & ...                            & ...                            & ...                            & ...                            & ...                            \\
IRAS01475-0740                 & ...                            & ...                            & ...                            & ...                            & ...                            & ...                            \\
IRAS08572+3915                 & ...                            & ...                            & ...                            & ...                            & ...                            & ...                            \\
IRAS09104+4109                 & ...                            & ...                            & ...                            & ...                            & ...                            & ...                            \\
IRAS10214+4724                 & ...                            & ...                            & ...                            & ...                            & ...                            & ...                            \\
IRAS13197-1627                 & ...                            & ...                            & ...                            & ...                            & $14.216\pm^{0.116}_{0.045}$    & $<4.3$                         \\
IRAS15480-0344                 & ...                            & ...                            & ...                            & ...                            & ...                            & ...                            \\
MCG-5-23-16                    & ...                            & ...                            & ...                            & ...                            & ...                            & ...                            \\
MRK3                           & ...                            & ...                            & $16.747\pm^{0.042}_{0.051}$    & $2.2\pm^{1.9}_{1.2}$           & $14.166\pm^{0.069}_{0.052}$    & $<4.8$                         \\
MRK6                           & ...                            & ...                            & ...                            & ...                            & ...                            & ...                            \\
MRK331                         & ...                            & ...                            & ...                            & ...                            & ...                            & ...                            \\
MRK348                         & ...                            & ...                            & ...                            & ...                            & ...                            & ...                            \\
MRK612                         & ...                            & ...                            & ...                            & ...                            & ...                            & ...                            \\
MRK744                         & ...                            & ...                            & ...                            & ...                            & ...                            & ...                            \\
MRK993                         & ...                            & ...                            & ...                            & ...                            & ...                            & ...                            \\
MRK1152                        & ...                            & ...                            & ...                            & ...                            & ...                            & $<100.7$                       \\
NGC1068                        & $31.486\pm^{0.034}_{0.035}$    & $1.6\pm^{0.2}_{0.1}$           & $16.817\pm^{0.044}_{0.035}$    & $8.9\pm^{0.5}_{0.8}$           & $14.199\pm^{0.035}_{0.035}$    & $2.7\pm^{1.1}_{0.8}$           \\
NGC1365                        & $31.511\pm^{0.159}_{0.085}$    & $1.6\pm^{0.2}_{0.1}$           & $16.775\pm^{0.054}_{0.046}$    & $2.3\pm^{2.0}_{1.0}$           & $14.197\pm^{0.044}_{0.044}$    & $<1.9$                         \\
NGC1386                        & ...                            & ...                            & $16.778\pm^{0.072}_{0.074}$    & ...                            & ...                            & ...                            \\
NGC1410                        & ...                            & ...                            & ...                            & ...                            & ...                            & ...                            \\
NGC1614                        & ...                            & ...                            & ...                            & ...                            & ...                            & ...                            \\
NGC2110                        & ...                            & ...                            & ...                            & ...                            & $14.158\pm^{0.053}_{0.039}$    & $<3.2$                         \\
NGC2273                        & ...                            & ...                            & ...                            & ...                            & ...                            & ...                            \\
NGC2623                        & ...                            & ...                            & ...                            & ...                            & ...                            & ...                            \\
NGC2992                        & ...                            & ...                            & ...                            & ...                            & ...                            & ...                            \\
NGC34                          & ...                            & ...                            & ...                            & ...                            & ...                            & ...                            \\
NGC3982                        & ...                            & ...                            & ...                            & ...                            & ...                            & ...                            \\
NGC4138                        & ...                            & ...                            & $16.775\pm^{0.056}_{0.081}$    & $<2.2$                         & ...                            & ...                            \\
NGC4151                        & $31.490\pm^{0.098}_{0.055}$    & $2.2\pm^{1.1}_{0.7}$           & $16.730\pm^{0.035}_{0.041}$    & $2.1\pm^{0.7}_{0.5}$           & $14.210\pm^{0.043}_{0.052}$    & $4.0\pm^{3.8}_{2.1}$           \\
NGC4168                        & ...                            & ...                            & ...                            & ...                            & ...                            & ...                            \\
NGC424                         & ...                            & ...                            & ...                            & ...                            & ...                            & ...                            \\
NGC4258                        & ...                            & ...                            & $16.929\pm^{0.051}_{0.051}$    & $9.5\pm^{1.4}_{2.1}$           & ...                            & $12.4\pm^{8.2}_{7.2}$          \\
NGC4303                        & ...                            & ...                            & $17.033\pm^{0.138}_{0.136}$    & $<13.9$                        & $14.197\pm^{0.061}_{0.072}$    & ...                            \\
NGC4395                        & ...                            & ...                            & ...                            & ...                            & $14.453\pm^{0.212}_{0.207}$    & $<66.6$                        \\
NGC4472                        & ...                            & ...                            & ...                            & ...                            & $14.234\pm^{0.215}_{0.080}$    & $7.0\pm^{10.4}_{4.7}$          \\
NGC4477                        & ...                            & ...                            & $17.041\pm^{0.126}_{0.125}$    & $<7.3$                         & ...                            & ...                            \\
NGC449                         & ...                            & ...                            & ...                            & ...                            & ...                            & ...                            \\
NGC4507                        & ...                            & ...                            & $16.740\pm^{0.049}_{0.052}$    & $2.3\pm^{2.2}_{1.1}$           & ...                            & $8.8\pm^{13.5}_{4.0}$          \\
NGC4565                        & ...                            & ...                            & ...                            & ...                            & ...                            & ...                            \\
NGC4639                        & ...                            & ...                            & ...                            & ...                            & ...                            & ...                            \\
NGC4725                        & ...                            & ...                            & ...                            & ...                            & ...                            & ...                            \\
NGC4945                        & ...                            & ...                            & $16.538\pm^{0.061}_{0.046}$    & $<5.0$                         & $14.232\pm^{0.251}_{0.244}$    & $<33.2$                        \\
NGC4968                        & ...                            & ...                            & ...                            & ...                            & ...                            & ...                            \\
NGC5033                        & $31.626\pm^{0.190}_{0.176}$    & $<6.4$                         & ...                            & ...                            & ...                            & ...                            \\
NGC5252                        & ...                            & ...                            & ...                            & ...                            & ...                            & ...                            \\
NGC526A                        & ...                            & ...                            & ...                            & ...                            & ...                            & ...                            \\
NGC5273                        & ...                            & ...                            & ...                            & ...                            & ...                            & ...                            \\
NGC5506                        & $31.630\pm^{0.091}_{0.081}$    & $<1.1$                         & $16.775\pm^{0.059}_{0.080}$    & $<5.6$                         & ...                            & ...                            \\
NGC5643                        & ...                            & ...                            & ...                            & $<22.0$                        & ...                            & ...                            \\
NGC591                         & ...                            & ...                            & ...                            & ...                            & ...                            & ...                            \\
NGC6552                        & ...                            & ...                            & ...                            & ...                            & ...                            & ...                            \\
NGC7172                        & ...                            & ...                            & ...                            & ...                            & ...                            & ...                            \\
NGC7212                        & ...                            & ...                            & ...                            & ...                            & ...                            & ...                            \\
NGC7314                        & ...                            & ...                            & ...                            & ...                            & ...                            & ...                            \\
NGC7479                        & ...                            & ...                            & ...                            & ...                            & ...                            & ...                            \\
NGC7582                        & ...                            & ...                            & $16.709\pm^{0.123}_{0.072}$    & $<6.3$                         & ...                            & ...                            \\
NGC7674                        & ...                            & ...                            & ...                            & ...                            & ...                            & ...                            \\
UGC1214                        & ...                            & ...                            & $16.897\pm^{0.271}_{0.263}$    & $3.5\pm^{5.3}_{3.3}$           & ...                            & $4.8\pm^{3.1}_{3.1}$           \\
UGC2456                        & ...                            & ...                            & ...                            & ...                            & $14.236\pm^{0.061}_{0.051}$    & $<5.2$                         \\
UGC2608                        & ...                            & ...                            & ...                            & ...                            & ...                            & ...                            \\
UGC4203                        & ...                            & ...                            & ...                            & ...                            & ...                            & ...                            \\
UGC6527                        & ...                            & ...                            & ...                            & ...                            & ...                            & ...                            \\
UGC8621                        & ...                            & ...                            & ...                            & ...                            & ...                            & ...                            \\
UM625                          & ...                            & ...                            & ...                            & ...                            & ...                            & ...                            \\
\hline \hline                                                                   
\end{tabular}                                                                   
\end{tiny}                                                                      
\caption{Best-fit wavelengths and widths for the RRCs discussed in this         
 {\it paper}. {\it Dots} indicates either lines not detected, or                
lines for which no constraints can be obtained (the luminosities in Tab.~\ref{tabidl5}              
are calculated in this case assuming the                                        
laboratory energies.)}                                                          
\label{tabidl4}                                                                 
\end{table*}
%------------------------- Table 4 IDL
%------------------------- Table 5 IDL
\begin{table*}                                                                  
\begin{tiny}                                                                    
\begin{tabular}{lccc} \hline \hline                                             
Source & C{\sc v} RRC & O{\sc vii} RRC & O{\sc viii} RRC  \\                    
 & $10^{-5}$~ph~cm$^{-2}$~s$^{-1}$ & $10^{-5}$~ph~cm$^{-2}$~s$^{-1}$ & $10^{-5}$~ph~cm$^{-2}$~s$^{-1}$  \\ \hline    
Circinus~Galaxy             & ...                                      & $2.4\pm^{19.3}_{2.3}$                    & $2.6\pm^{1.7}_{1.5}$       \\
ESO509-G66                 & ...                                      & ...                                      & ...                        \\
IC2560                     & ...                                      & $0.3\pm^{0.2}_{0.2}$                     & ...                        \\
IC4395                     & ...                                      & ...                                      & ...                        \\
IC4995                     & ...                                      & ...                                      & ...                        \\
IIIZW035                   & ...                                      & ...                                      & ...                        \\
IRAS01475-0740             & ...                                      & ...                                      & ...                        \\
IRAS08572+3915             & ...                                      & ...                                      & ...                        \\
IRAS09104+4109             & ...                                      & ...                                      & ...                        \\
IRAS10214+4724            & ...                                      & ...                                      & ...                        \\
IRAS13197-1627             & ...                                     & ...                                      & $0.6\pm^{0.5}_{0.3}$       \\
IRAS15480-0344             & ...                                      & ...                                      & ...                        \\
MCG-5-23-16                & ...                                      & ...                                      & ...                        \\
MRK3                 & ...                                      & $1.6\pm^{0.7}_{0.4}$                     & $0.9\pm^{0.5}_{0.3}$       \\
MRK6                 & ...                                      & ...                                      & ...                        \\
MRK331                     & ...                                      & ...                                      & ...                        \\
MRK348                     & ...                                      & ...                                      & ...                        \\
MRK612                     & ...                                      & ...                                      & ...                        \\
MRK744                     & ...                                      & ...                                      & ...                        \\
MRK993                     & ...                                      & ...                                      & ...                        \\
MRK1152                    & ...                                      & ...                                      & $13.6\pm^{27.4}_{12.0}$    \\
NGC1068                    & $67.2\pm^{10.1}_{7.7}$                   & $65.9\pm^{2.1}_{9.1}$                    & $10.1\pm^{2.7}_{1.9}$      \\
NGC1365                    & $0.8\pm^{0.5}_{0.5}$                     & $1.1\pm^{0.5}_{0.3}$                     & $0.5\pm^{0.2}_{0.2}$       \\
NGC1386                    & ...                                      & $0.3\pm^{1.5}_{0.3}$                     & ...                        \\
NGC1410                    & ...                                      & ...                                      & ...                        \\
NGC1614                    & ...                                      & ...                                      & ...                        \\
NGC2110                    & ...                                      & ...                                      & $1.4\pm^{0.6}_{0.5}$       \\
NGC2273                    & ...                                      & ...                                      & ...                        \\
NGC2623                    & ...                                      & ...                                      & ...                        \\
NGC2992                    & ...                                      & ...                                      & ...                        \\
NGC34                      & ...                                      & ...                                      & ...                        \\
NGC3982                    & $1.1\pm^{1.4}_{0.8}$                     & ...                                      & ...                        \\
NGC4138                    & ...                                      & $0.4\pm^{0.4}_{0.3}$                     & ...                        \\
NGC4151                    & $13.9\pm^{8.5}_{7.1}$                    & $5.7\pm^{1.2}_{1.0}$                     & $4.3\pm^{6.4}_{1.6}$       \\
NGC4168                    & ...                                      & ...                                      & ...                        \\
NGC424                     & ...                                      & ...                                      & ...                        \\
NGC4258                    & ...                                      & $7.6\pm^{0.8}_{2.4}$                     & ...                        \\
NGC4303                    & ...                                      & $1.4\pm^{1.5}_{1.0}$                     & $0.3\pm^{0.3}_{0.2}$       \\
NGC4395                    & ...                                      & ...                                      & $0.4\pm^{1.8}_{0.3}$       \\
NGC4472                    & ...                                      & ...                                      & $22.3\pm^{8.6}_{16.0}$     \\
NGC4477                    & ...                                      & $1.6\pm^{1.8}_{1.0}$                     & ...                        \\
NGC449                     & ...                                      & ...                                      & ...                        \\
NGC4507                    & ...                                      & $1.7\pm^{0.6}_{0.6}$                     & ...                        \\
NGC4565                    & ...                                      & ...                                      & ...                        \\
NGC4639                    & ...                                      & ...                                      & ...                        \\
NGC4725                    & ...                                      & $0.3\pm^{0.3}_{0.2}$                     & ...                        \\
NGC4945                    & ...                                      & ...                                      & ...                        \\
NGC4968                    & ...                                      & ...                                      & ...                        \\
NGC5033                    & $1.4\pm^{2.5}_{1.2}$                     & ...                                      & ...                        \\
NGC5252                    & ...                                      & ...                                      & ...                        \\
NGC526A                    & ...                                      & ...                                      & ...                        \\
NGC5273                    & ...                                      & ...                                      & ...                        \\
NGC5506                    & $1.4\pm^{1.0}_{1.1}$                     & $0.9\pm^{0.4}_{0.5}$                     & ...                        \\
NGC5643                    & ...                                      & $2.0\pm^{3.1}_{1.7}$                     & ...                        \\
NGC591                     & ...                                      & ...                                      & ...                        \\
NGC6552                    & ...                                      & ...                                      & ...                        \\
NGC7172                    & ...                                      & ...                                      & ...                        \\
NGC7212                    & ...                                      & ...                                      & ...                        \\
NGC7314                    & ...                                      & ...                                      & ...                        \\
NGC7479                    & ...                                      & ...                                      & ...                        \\
NGC7582                    & ...                                      & $0.5\pm^{0.4}_{0.3}$                     & $0.7\pm^{0.6}_{0.4}$       \\
NGC7674                    & ...                                      & ...                                      & ...                        \\
UGC1214                    & ...                                      & $1.5\pm^{1.5}_{0.9}$                     & ...                        \\
UGC2456                    & ...                                      & ...                                      & $1.2\pm^{1.1}_{0.8}$       \\
UGC2608                    & ...                                      & ...                                      & ...                        \\
UGC4203                    & ...                                      & ...                                      & ...                        \\
UGC6527                    & ...                                      & ...                                      & ...                        \\
UGC8621                    & ...                                      & ...                                      & ...                        \\
UM625                      & ...                                      & ...                                      & ...                        \\
\hline \hline                                                                   
\end{tabular}                                                                   
\end{tiny}                                                                      
\caption{Best-fit fluxes for the RRCs discussed in this {\it paper}.            
{\it Dots} indicate lines not detected.}                                       
\label{tabidl5}                                                                 
\end{table*}                             
%------------------------- Table 5 IDL

The comparison between the best-fit centroid energy of
detected lines in {\it CIELO-AGN} and the laboratory energies is an instructive
exercise on the reliability of the catalog measurements. At the 2$\sigma$
level, the number of He- and H-like Oxygen
lines (out the total detected ones) which are inconsistent
with the laboratory energies are: 0/25, 2/20, 2/30 and 5/22
for O{\sc vii} He-$\alpha$, O{\sc vii} He-$\beta$,
O{\sc viii} Ly-${\alpha}$ and O{\sc viii} Ly-${\beta}$,
respectively. The shifted measurements in the last case
cluster around 783~eV (4 out of 5), where a strong Fe{\sc xviii}~3s2p
transition is located ($\lambda = 15.83$~\AA), and are
therefore probably due to a mis-identification. If we
remove these suspicious cases, the ratio between
the He-${\beta}$ and the $f$ transition
intensity in the
remaining objects is still inconsistent with the expectation
of pure photoionization: $0.26 \pm 0.04$
(cf. Tab.~\ref{tab1}). The situation for the C{\sc vi} lines is
more controversial. The centroid energies
of 6 out of 23 Ly-${\alpha}$ and
5 out of 14 Ly-${\beta}$ measurements are inconsistent at
the 2$\sigma$ level with the laboratory energies. There
is no obvious explanation for these systematic discrepancies.
No other potentially contaminating transition
is present, which could confuse the identification of
the C{\sc vi} lines.
Moreover, the shifted measurements
do not cluster in any well defined spectral region.
Uncertainties on the aspect solutions can be
as high as 30\AA\ in this wavelength range, still insufficient
to explain the observed differences.

A similar analysis on the centroid energies of the
RRC features shows that only 3 measurements (out of 16) of the O{\sc vii}
RRC are inconsistent with the laboratory energy (at the 2$\sigma$ level;
none for the C{\sc v} or O{\sc viii} RRCs).
In this case, as discussed in Sect.~3.1, the discrepancy is
probably due to confusion with the nearby Fe{\sc xvii} triplet.

\section*{Acknowledgments}

This paper is based on observations obtained with XMM-Newton, an ESA
science mission with instruments and contributions directly funded by
ESA Member States and the USA (NASA). This research has made use of
data obtained through the High Energy Astrophysics Science Archive
Research Center On-line Service, provided by the NASA/Goddard Space
Flight Center and of the NASA/IPAC Extragalactic Database (NED) which
is operated by the Jet Propulsion Laboratory, California Institute of
Technology, under contract with the National Aeronautics and Space
Administration. Discussions with A.Pollock and J.Sanz allowed us
to better understand the properties of high-resolution
X-ray spectra of active stars.
We are deeply grateful to Dr.A.Kinkhabwala,
for providing us with an updated version of his
{\tt photoion} code, as well as for
several enlightening and encouraging suggestions and comments.
Comments and encouragement by an anonymous referee, which
significantly improved the quality of this paper, are
warmly acknowledged.

{}

\end{document}